\documentclass[runningheads]{llncs}

\usepackage[utf8]{inputenc}
\usepackage{url}
\usepackage[misc]{ifsym}

\usepackage{booktabs}

\usepackage{tikz}
\usetikzlibrary{automata, positioning, arrows}

\usepackage{amssymb}

\graphicspath{{./figures/}}
\DeclareGraphicsExtensions{.pdf,.png}

\usepackage{color, colortbl}

\usepackage{breqn}

\newcolumntype{L}{>{\centering\arraybackslash}p{2cm}}
\newcolumntype{T}{>{\centering\arraybackslash}p{2.5cm}}
\newcolumntype{D}{>{\arraybackslash}p{4.5cm}}
\newcolumntype{E}{>{\arraybackslash}p{9cm}}

\begin{document}

\title{The uncertainty of Side-Channel Analysis}
\subtitle{A way to leverage from heuristics}
\titlerunning{The uncertainty of Side-Channel Analysis: A way to leverage from heuristics}

\author{Unai Rioja\inst{1,2,}\textsuperscript{(\Letter)}\thanks{Both authors contributed equally to this research.} \and
Servio Paguada \inst{1,2,}\textsuperscript{(\Letter)}\textsuperscript{$\star$ } \and
Lejla Batina\inst{1,}\textsuperscript{(\Letter)} \and
Igor Armendariz\inst{2,}\textsuperscript{(\Letter)}}

\authorrunning{Rioja and Paguada et al.}

\institute{Digital Security Group, Radboud University Nijmegen, The Netherlands \\
\email{lejla@cs.ru.nl} \and
Ikerlan Technology Research Centre, Arrasate-Mondragón, Gipuzkoa, Spain \\
\email{\{urioja,slpaguada,iarmendariz\}@ikerlan.es}}

\maketitle             

\begin{abstract}
 Performing a comprehensive side-channel analysis evaluation of small embedded devices is a process known for its variability and complexity. In real-world experimental setups, the results are largely influenced by a huge amount of parameters that are not easily adjusted without trial and error and are heavily relying on the experience of professional security analysts. In this paper, we advocate the use of an existing statistical methodology called Six Sigma (6$\sigma$) for side-channel analysis optimization for this purpose. This well-known methodology is commonly used in other industrial fields, such as production and quality engineering, to reduce the variability of industrial processes. We propose a customized Six Sigma methodology, which enables even a less-experienced security analysis to select optimal values for the different variables that are critical for the side-channel analysis procedure. Moreover, we show how our methodology helps in improving different phases in the side-channel analysis process.
\end{abstract}


\section{Introduction}
\label{sec.Intro}
The process of obtaining data for Side-Channel Analysis (SCA) is a complex procedure in which not only the acquisition of thousands or even millions of power or EM traces is needed, but also the usage of signal processing techniques in combination with advanced statistical and mathematical tools is most of the times mandatory. In every step of this path, many decisions have to be taken most commonly based on the know-how of the people who have dealt with this kind of issues in the past. Obtaining data for SCA (and performing the SCA itself) is an unpredictable process in which decisions are usually taken via repeated ``trial and error''. Moreover, building a proper experimental setup is often a non-repetitive task, namely what is appropriate for one device can be inefficient for others. In common scenarios, security analysts deal with lots of parameters to tune and choices to make. Thus, no wonder that at the end of the day, when good results arise after a huge amount of changes applied on the spot, analysts often do not know exactly what decision or what parameter caused the improvement of the results. In practice, as the Pareto principle postulates \cite{Pareto}, there are a few parameters that have a strong influence in the results (vital few) against lots of parameters whose impact in the results is negligible (trivial many). It is not easy to find one parameter (or a group of parameters) that causes the biggest effect on the experiment. Therefore, it is important to use a suitable methodology to perform an experimentation process with those characteristics.

Six Sigma is a well-known statistical methodology targeted to improve industrial processes (production and quality engineering) by reducing its variability. To the best of our knowledge, it has not been used before to reduce the uncertainty of an SCA process. We develop a customized version of Six Sigma to make it fit to the SCA requirements and to be able to use it in optimizing each one of the SCA phases e.g. acquisition, leakage assessment and attacking phase. After applying the methodology, we were able to select the best values for the different parameters analyzed within distinct SCA scenarios. In addition, it helped us in finding the most relevant parameters for the analyses results.

\noindent
\textit{Problem statement:} Systematically keeping track of the parameters in an evaluation or an attack scenario is not trivial, and also not feasible without a well-defined procedure. The current state-of-the-art approach is often experimental and founded on previous experiences. However, this strategy does not give much insight in the choice of the parameters that could have the most impact on the results.

\noindent
\textit{Our contribution:} In this paper, we do not claim having developed a new methodology for leakage assessments nor for performing new attacks on a device. Instead, we present an approach that not only complements leakage assessment techniques, but also enhances attacking scenario setups. This is done without the need for any new resources (in time or power). We claim that, by using this new methodology, one could reduce the uncertainty associated with those, often cumbersome, techniques. The goal is to help lab technicians (or even product developers), who may not have a deep knowledge of all the statistical and signal processing concepts involved in these methods, to perform a sound side-channel evaluation and interpret the results properly. In the case of a more experienced evaluator, to use the methodology when he faces a new set of devices or data sets could lead him to discover the best set of variables, that impact the most with regard to the new task just by conducting it at least one time.

The rest of the paper is organized as follows: Sect. \ref{sec.6sigma} introduces main concepts of the Six Sigma methodology, explains its main steps and connects it to the practical use cases. The goal of our use cases is to show how Six Sigma can help in optimizing each one of the side channel analysis phases, including acquisition setup optimization (Sect. \ref{UC1}), attack optimization (Sect. \ref{UC2}) and leakage assessment optimization (Sections \ref{UC3} and \ref{UC4}). Finally, Sect. \ref{CONCLUSION}, concludes the paper.

\section{Six Sigma Methodology}
\label{sec.6sigma}

The Six Sigma (6$\sigma$) methodology was created in 1986 by Bill Smith, while working as an engineer at
Motorola, as the company that registered the term as its trademark in 1993 \cite{6sigma}. The primary objective was to minimize the variability of the output 
of a process. To achieve this, different empirical quality management methods, along with statistical methods are used. In this improvement process some steps have to be repeated until the main goal is reached. 

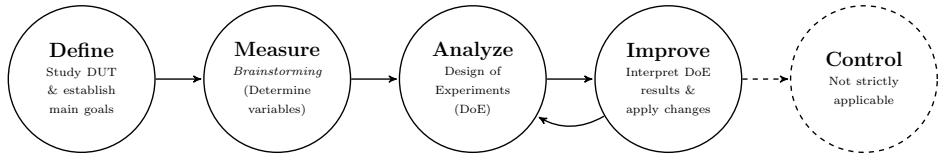
\begin{figure}[htbp]
\begin{center}
	\scalebox{0.65}{\begin{tikzpicture}[->,>=stealth',shorten >=1pt,auto,node distance=4cm,
	thick,main node/.style={circle,draw},minimum size=3cm]
	
	\node[main node] (0) [text width=2cm, align=center]{\large \textbf{Define} \\ \scriptsize Study DUT \&  establish main goals};
	\node[main node] (1) [right of=0, text width=2cm, align=center] {\large \textbf{Measure} \\ \scriptsize \textit{Brainstorming} (Determine variables)};
	\node[main node] (2) [right of=1, text width=2cm, align=center] {\large \textbf{Analyze} \\ \scriptsize Design of Experiments (DoE)};
	\node[main node] (3) [right of=2, text width=2cm, align=center] {\large \textbf{Improve} \\ \scriptsize Interpret DoE results \& apply changes};
	\node[main node,dashed] (4) [right of=3, text width=2cm, align=center] {\large \textbf{Control} \\ \scriptsize Not strictly applicable};
	
	\path[every node/.style={font=\sffamily\small}]
	(0) edge[above] (1)
	(1) edge[above] (2)
	(2) edge[above] (3)
	(3) edge[bend left, below] (2)
	(3) edge[dashed] (4);
	\end{tikzpicture}}
\end{center}
\caption{Customized Six Sigma methodology steps.}
\label{fig:six_sigma_steps}
\end{figure}

Six Sigma involves two main methodologies implicitly, which are the basis for the 
process management and optimization, and also for the guidelines proposed in this 
document. The methodologies are:  \textbf{Define}-\textbf{Measure}-\textbf{Analyze}-\textbf{Improve}-\textbf{Control} (DMAIC) and \textbf{Define}-\textbf{Measure}-\textbf{Analyze}-\textbf{Design}-\textbf{Verify} (DMADV). The former aims at improving an existing process, whilst the latter aims at designing a new process. Both are based on the Deming’s Plan-Do-Check-Act Cycle \cite{Srinivas2018,cheng2012}. Although those two methodologies are similar, we focus on DMAIC, because our aim is to improve the selection of parameters for an SCA process, and not to design the process itself. The steps of the methodology are shown in Figure \ref{fig:six_sigma_steps}, pointing how they fit in an SCA evaluation use case (a detailed explanation can be found in Sect. \ref{UC1}).

\begin{itemize}
\item \textbf{Define} the system. Here, as the system's inputs we envision the client's requirements and the goal of the project i.e. evaluation. In our case, the ``Define phase'' is basically the study of the Device Under Test (DUT). When a ``black-box'' evaluation is performed, one deals with a device with almost no information about its internals. Nevertheless, in a ``white-box'' evaluation there are also some variables with less certainty e.g., working frequency, algorithm implementation details, location of the cryptographic operations etc. Depending on how evaluators define their goal, the ``uncertain'' variables are revealed systematically by the end of the whole process. In this step, one defines not only the main goal(s) but also the OK-criterion. The proper definition of those helps to minimize the uncertainties. We interpret this OK-criterion as the quantification of the goal we want to achieve. In other words, this implies the definition of a factor that allows us to decide if the experiments are conclusive or not. A more detailed explanation can be found in the ``Define'' section of use cases, i.e. Sect. \ref{UC1.DEF}, \ref{UC2.DEF}, and \ref{UC3.DEF}.

\item \textbf{Measure} the current process setup. To characterize the current state of the process, one collects its parameters and outputs. In our case, a few preliminary acquisitions of side-channel signals should be done, to be used as a baseline for the results that are meant to be improved. The objective is to define the variables/parameters we are going to study. We need to define the system's variables that could affect the quality of the experiments (e.g. number of traces, sampling frequency, filtering). For those variables, one prescribes a minimum and a maximum value. Commonly, those values depend on a huge number of factors, so consequently, it could be necessary to perform several preliminary measurements to establish the right values.	Once the variables are defined and bounded, we must choose three of them; the three variables that are most likely to affect the results of the experiment. The rest of them will be fixed to some constant values (e.g. their minimum and maximum). To illustrate this, in the ``Measure'' section of our use cases, (i.e. Sect. \ref{UC1.MEAS}, \ref{UC2.MEAS}, and \ref{UC3.MEAS}), the reader can find Tables \ref{ParameterTable1}, \ref{ParameterTable2}, and \ref{ParameterTable3} with ``Defined variables'', with all the considered variables, and Tables \ref{ParameterSelection1}, \ref{ParameterSelection2}, and \ref{ParameterSelection3} with the 3 chosen as ``Working variables'' along with its minimum and maximum values for each respective case.
	
\item \textbf{Analyze} the data obtained from the process, and determine its relationships with the problem. This step consists of experimentation i.e., crafting an experiment or Design of Experiments (DoE). DoE is a branch of applied statistics, which is responsible for evaluating the factors (or variables) that influence a parameter or group of parameters. Note that in this paper we do not explain DoE in detail, but we refer interested readers to \cite{Montgomery:2019:DAE}, and \cite{DOE}. The objective is to quantify which variables have more influence over the experiment and adjust them to the proper values. To do that, a DoE with the 3 selected working variables to perform 8 experiments is chosen. The output of it gives the coefficient to each variable, which tells us if the effect is positive or negative (improves or not the result of the experiment) and how strong each one in comparison with the others is. In the ``Analyze section'' of use cases (Sect.~\ref{UC1.ANA}, Sect.~\ref{UC2.ANA}, and Sect.~\ref{UC3.ANA}) the output of each DoE can be observed with more detail.

\item \textbf{Improve} the current process using the analysis of the root 
causes done in the previous step to identify, test and implement a solution for the problems that appeared. In this customized Six Sigma, the step consists of the analysis of the experimental design’s results. Here one adjusts the identified working variables that have more influence over the experiments. Afterwards, one performs many rounds of the experiment (at least 2 times), to ensure that the results are not altered due to any failure. If this is not the case, a new setup is designed considering those results. If the results are not good enough (even after the 8 experiments), the process should be repeated from the previous step, considering to change the selected variables or adjust their minimum and maximum values. This is considered as one iteration. The idea is to perform several iterations between these two last steps until the main objective is reached. Practical examples of this analysis of the results and the readjustment of the DoE variables are shown in the ``Improve'' section of our use cases (Sect.~\ref{UC1.IMPRO}, Sect.~\ref{UC2.IMPRO}, and Sect.~\ref{UC3.IMPRO}).

\item \textbf{Control} the newly improved process to correct any undesired deviations of it. Repeat the steps until obtaining the desired quality level. This step does not strictly apply to our problem, but it can be understood as the action of taking notes of the results to apply in futures experiments. 
\end{itemize}

As mentioned above, the proposed methodology can be used to optimize acquisition, leakage assessment, attack evaluation, and parameter selection. Below we describe those four use cases using this customized Six Sigma methodology in different scenarios. The first use case is also taken as an explanatory one, which means that it is explained in more detail than the rest. 

\section{Use case 1: Improving the acquisition phase}
\label{UC1}
In this section the proposed method is presented and explained step by step, giving examples for the procedures done with our experimental setup to optimize the acquisition process over a real device. First of all, we briefly describe the presented use case, then we divide the section into 5 steps following the DMAIC scheme. 

\subsection{Use case description}
\label{UC1.UCD}
The target is an external I2C device (slave). We are storing data (8-bit values) in an external I2C memory using the STM32F4 32-bit microcontroller (master). During the storage operation, the I2C device's power consumption is being measured with a current probe. As evaluators, we would like to know whether it is possible to find dependency between the stored data and the power consumption, allowing an attacker to perform SCA. We use a proper experimental setup to do that and we apply our customized Six Sigma methodology, to optimize the acquisition setup. 

After a few preliminary acquisitions (taken with the setup shown in Table \ref{ToolList}) we did not learn much about the point in time when the data was being stored, and we were not able to obtain any correlation between the stored data and the power traces. Then, we used our customized Six Sigma methodology to improve the acquisition setup and to obtain a significant correlation spike, identifying the exact point of time in which the 8-bit value was being stored in memory (Figure \ref{fig:icorrelationFound}).

\subsection{Define}
\label{UC1.DEF}
First, the \textit{Device-Under-Test} (DUT) should be studied by an analyst to get familiar with it. Although this process should be always done prior to any SCA experiment, it is crucial for the ``black-box'' testing use cases, since its results might be the only source of information about the DUT. From this analysis, enough 
information should be gathered to define the following: the acquisition setup's requirements, the goal of analyzing the acquired data, and the \textit{OK-criterion}. We propose several steps for this preliminary analysis. Note that this use case is for handling devices whose internals are mostly unknown, while examining side-channel leakage from them (e.g., ``black-box'' testing or an attack scenario). Most of these steps can be avoided if there is enough information about the internal behavior (such as in a ``white-box'' testing scenario in which a complete control of the device is assumed).

\begin{itemize}
    \item[a.] Define the basis of the unknown system, assuming some basic information is known. Usually, there are some known characteristics like its purpose, manufacturer, inputs and outputs, how much power (externally) it consumes, etc. Everything that is clearly known (and not guessed), should be written down as a list, as it is a starting point for the questions about characteristics of the system that are (still) unknown. For instance: Does the manufacturer uses standard architectures? What are the operations that the power consumption depends on? What operations are performed with the input data? 
    Part of the answers on those questions can determine some of the initial parameters of the acquisition setup, like the voltage scale in what the measurements should be done.
    
	\item[b.] Analyse all the official (device data sheet) and unofficial documentation that can be obtained, trying to infer the details of the internal architecture that are not explicitly mentioned by the manufacturer.
	
	\item[c.] Apply non-invasive analysis techniques on it, like measuring the voltage and performing a Fourier analysis, to learn the working frequency. There are also invasive or more destructive techniques possible but they are out of scope here, since we focus on passive analysis.
	
	\item[d.] Finally, the experts' knowledge on similar devices can be helpful, but it is not mandatory.
\end{itemize}
	
After the analysis, one has to summarize the gathered information, enhancing the basic knowledge about the system with the new information obtained through the comparison of documentation, the non-invasive analysis and the opinion of the experts. In our use case, first the data sheet provided by the manufacturer was read and thoroughly analyzed. After that, the process of getting system's details mainly consisted of voltage measuring and performing a frequency analysis, from which a few different frequencies were identified. Both processes were accomplished by a person with some expertise in the field, and the results were discussed with the expert in electronic devices of the team.

Once the DUT has been thoroughly studied, we can use this information to define the requirements of the acquisition setup, the main goal of the experiment and the \textit{OK-criterion}:\\

\noindent
\textbf{Requirements of the setup:} We interpret this as the initial parameters of the acquisition setup, which are not supposed to change during all the steps of the process. For instance, these parameters could be (but not strictly): operational parameters of the target, acquisition/analysis tools, physical properties studied (e.g., power consumption, electromagnetic radiation, and timing), type of the SCA technique implemented, etc. If, while applying the process something suggests to change this data, this should be taken into account, continuing the process until the end, and adding those changes in a future iteration. The setup requirements for our use case are given in Table ~\ref{setup_goal}.\\

\noindent
\textbf{Main goal:} Here we define the goal that we want to achieve. In our case, the experiment was to obtain and visualize a dependency between the stored 8-bit values and the power consumption.\\
    
\noindent
\textbf{\textit{OK-criterion}:} This is the quantification of the goal, a factor that tells us whether the experiments are conclusive or not. During the experiments one tries to tune several parameters obtaining different results through iterations, so the \textit{OK-criterion} indicates when to stop the Six Sigma performance. Also, it tells when we have reached our objectives. In this use case, the \textit{OK-criterion} (Table \ref{setup_goal}) is to obtain a significant correlation spike (in comparison to the correlation obtained in non-leaking parts of the signal), indicating that in fact there is a data-dependency with the power consumption of the device. Moreover, we can compute the \textbf{Confidence intervals} for a sample (Pearson) coefficient value \cite{personGREENspringer} to establish a concrete threshold. With a sample size of 1k and an observed correlation of approximately $\pm$0.05 (in the non leaking parts), our confidence interval (99.99\%) is \textbf{from -0.1705 to 0.1705} (considering positive and negative correlation values). This means that if the observed correlation is out of that range, we can assume that there exists a significant statistical difference.

\begin{table}[htbp] 
		\centering
		\caption{\label{setup_goal} Goal and \textit{OK-criterion}}
		\scalebox{1}{\begin{tabular}{ T  E }
            \toprule
			Goal & Find correlation between the power traces and the stored data \\
		    \hline
			Requirements & \begin{minipage}[t]{0.65\textwidth} \begin{itemize} 
						\item Same setup and device operational parameters fixed an constant in each experiment
						\item Oscilloscope: LeCroy Waverunner 9104 
						\item SW for the acquisition, signal processing and data 
						analysis 
						\item Current probe
				\end{itemize} \end{minipage}\\ 
		    \hline
				\textit{OK-criterion} & Significant correlation spike ($r$) [Where $ (r \leq -0.1705)  \cap (r \geq 0.1705)$]  \\
		    \bottomrule
		\end{tabular}}
	\end{table}

After having defined the basic requirements for the acquisition setup, the goal of acquiring the data, and the \textit{OK-criterion} to evaluate the improvement on the results the measurement step can be started.

\subsection{Measure}
\label{UC1.MEAS}

Following the requirements defined in the previous step, the usual measurement setup is parametrized, and a first preliminary round of acquisitions is started with this basic setup. This basic setup consists of the following tools configured as stated below (Table \ref{ToolList}).

\begin{table}[htbp]
	\centering
	\caption{Tools list and brief description.}
	\label{ToolList}
	\scalebox{1}{\begin{tabular}{TE}
			\toprule
			LeCroy Waverunner 9104 Oscilloscope & 
			\begin{minipage}[t]{0.65\textwidth}
		    \begin{itemize} 
					\item 2 channels (Power consumption \& triggering) 
					\item 20 GS/s (Single sample capture mode)
					\item I2C bus-based triggering
			\end{itemize}\end{minipage}\\
			
			\hline
			The DUT  & \begin{minipage}[t]{0.65\textwidth} \begin{itemize} 
					\item Handled by an STM32F411-DISCO developing board (I2C based communication)
					\item Power supplied through the STM32F411-DISCO developing board
			\end{itemize} \end{minipage}\\ 
			\hline
			PC  & \begin{minipage}[t]{0.65\textwidth} \begin{itemize}
					\item Communicates with the STM32F411-DISCO developing board
					\item Controls the oscilloscope
			\end{itemize} \end{minipage}\\ 
			\hline
			Current Probe  & \begin{minipage}[t]{0.65\textwidth} \begin{itemize} 
					\item Tektronix CT1 current probe
					\item Conected in series with the DUT power line
					\item Measure the power consumption
			\end{itemize} \end{minipage}\\ 
			\bottomrule
	\end{tabular}}
\end{table}

With this baseline setup, a round of random 8-bit values are sent to the DUT from the computer (through the STM32F4 developing board), while the oscilloscope triggers in the I2C clock line (SCL), just after finishing the communication, when the internal computation in the DUT is supposed to start. The main goal of this step is to define the system's variables that we are going to study. To ensure quality traces, the evaluator acquires a couple of hundreds of them, and in those he might also discover defects or inconvenient features. Accordingly, he/she might decide to tune further some parameters to improve the results. The parameters can be different in nature (environment, processing, measurement, etc.). Also, there are some parameters that can be considered in most of the cases (Lowpass filtering, number of traces, compression techniques, etc.) while other parameters will depend on the specific use case.

After prompting all the possible parameters that need further tuning, the expert should evaluate and order them by their potential being more significant for improving the quality of the traces. For each parameter a minimum and a maximum values have to be specified (some parameters will be boolean but others will have a range of possible values). Also, this list must be analyzed to avoid the selection of parameters that can be dependent on each other. Table \ref{ParameterTable1} gives the parameters with their descriptions and  ranges. The top three variables will be analyzed in the next step performing a DoE on them (Table \ref{ParameterSelection1}). The rest of the parameters have to be fixed in values between their minimum and maximum.

\begin{table}[!h]
	\centering
	\caption{Defined variables}
	\label{ParameterTable1}
	\scalebox{1}{\begin{tabular}{cLDLL}
			\toprule
			\textbf{Rank} & \textbf{Parameter} & \textbf{Description} & \textbf{Range} & \textbf{Fixed Value}\\
			\midrule
			1 & Point of Alignment & 
			Two interesting places in the traces to search for correlation are observed& 
			Align at the start vs at the end &
			\\
			2 & LowPass Filter (SW) &
			Filtering the signal or not would affect the quality of the collected traces. We should eliminate high frequency noise but without destroying the leakage. &
			Filtering vs No filtering &
			\\
			3 & Standard-ization &
			Removing the mean of the data set can help to improve the results. &
			Yes vs No &
			\\
		    4 & Nº of traces &
			The number of traces affects to the confidence of the results but increases the computational effort & 
			1k vs 100k &
			\textbf{1k}\\
			5 & Compress-ion &
			Compression can be used to reduce the dataset size and to improve the leakage (close points carry very similar information and noise can be reduced). Conversely, leakage can also be destroyed because of compression. &
		    compression vs no compression &
			\textbf{No compression}\\
			6 & Sampling frequency &
			A high sampling frequency will improve the quality of the traces but also increase the data size &
			50MHz vs 20GHz &
			\textbf{1 GHz} \\
			\bottomrule
	\end{tabular}}
\end{table}

\subsection{Analyze}
\label{UC1.ANA}
\begin{figure}[!h]
	\centering
	\scalebox{1}{\includegraphics[width=0.9\textwidth]{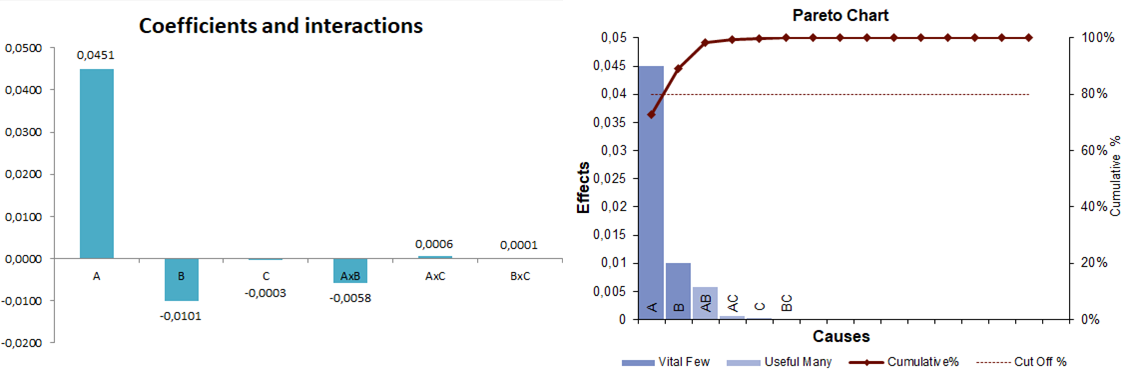}}
	\caption{\label{fig:iDOE_1_1_AVG_&_PARETO} Coefficients and interactions of the DoE (left) and Pareto chart (right)}
\end{figure}

From the ordered list, the three top parameters are chosen, and a simple Design of Experiments (DoE \cite{DOE}) process is carried out. In our case, we choose a two level DoE with 3 variables (called factors in experimental design) for its simplicity and reliability. Therefore, 3 variables are investigated in two levels by performing $2^3$ experiments as follows:

\begin{enumerate}
	\item Create an experiment matrix for all possible combinations of parameters
	\itemsep 0em 
	\item Select the two most suitable values (or minimum and maximum) for each selected parameter.
	\item Enter the selected parameter values in the matrix.
	\item Proceed to acquire the traces of the 8 experiments. 
	\item Process these traces in the selected SCA suite and write down the results.
	\item Calculate the \textit{Effect} and \textit{Coefficient} of each parameter with the following formula: 
	
	\begin{center}
	$ \textrm{\textit{Effect}} = \sum_{i=1}^4\textrm{\textit{Maximum}}_i - \sum_{i=1}^4\textrm{\textit{Minimum}}_i$
\end{center}
\begin{center}
	\begin{equation}\label{coef1} \textrm{\textit{Coefficient}} = c 
	=\frac{\textrm{\textit{Effect}}}{2}\end{equation}
\end{center}
	
	\item Calculate the \textit{Effect} and \textit{Coefficient}  of the interactions between the parameters with the following formula: 
	
	\begin{center}$ \textrm{\textit{Effect}} =  \sum_{i=1}^4positive Interaction_i -  
	\sum_{i=1}^4negative Interaction_i$ 
    \end{center}
    \begin{center}\begin{equation}\label{coef2} \textrm{\textit{Coefficient}} = c = 
    	\frac{\textrm{\textit{Effect}}}{2}\end{equation}
    \end{center}
    Note: the sign of the interaction is the product of the code for the parameter levels: $Minimum = -1$, $Maximum = 1$.
	
	\item Finally, calculate the results applying the DoE formula using the \textit{Coefficients} for each factor.
    
    \begin{center}
    $ DoE =  \sum_{i=1}^8\frac{R_i}{8} +  c_A*A + c_B*B +c_C*C + c_AB*AB+c_AC*AC+c_AC*AC $
    \end{center}

	\item Optionally, the results can be plotted with a Pareto Chart, for their better understanding (Fig. \ref{fig:iDOE_1_1_AVG_&_PARETO}, right).
	\item It is recommended to do more than one round of experiments with the same permutation list to compute the confidence of the results getting the variance $\sigma$ and indicating the error.
\end{enumerate}

\begin{table}[htbp]
	\centering
	\caption{Working variables and values}
	\label{ParameterSelection1}
	\scalebox{1}{\begin{tabular}{cccc}
		\toprule
		\textbf{ID} & \textbf{Parameter} &  \textbf{$1^{st}$ option (-)}& \textbf{ $ 2^{nd}$ option (+)}\\
		\midrule
		A & Point of alignment & Align at the start & Align at the end\\ 
		B & LowPass filter (Software) & false & true\\ 
		C & Standardization & false & true\\
		\bottomrule
	\end{tabular}}
\vspace{0.5cm}
		\centering
		\caption{Experiment definition and order}
		\label{ExperimentPermutations}
		\scalebox{1}{\begin{tabular}{ccccccc}
				\toprule
				\textbf{Experiment} &  \textbf{A}&  \textbf{Value} &  \textbf{B} & \textbf{Value} &  \textbf{C} &  \textbf{Value}\\
				\midrule
				1&-&start&-&false&-&false\\
				2&-&start&-&false&+&true\\
				3&-&start&+&true&-&false\\
				4&-&start&+&true&+&true\\
				5&+&end&-&false&-&false\\
				6&+&end&-&false&+&true\\
				7&+&end&+&true&-&false\\
				8&+&end&+&true&+&true\\
				\bottomrule
		\end{tabular}}
\end{table}

If the results of this process do not show a clear gain in any of the parameters, the next parameters in the ordered list are selected and the DoE applied again. In the case of our experimental setup, we were not able to find any correlation with the basic setup (Fig. \ref{fig:icorrelationFound}). Some misalignments and very noisy signals were detected, so the three parameters shown in Table \ref{ParameterSelection1} were selected. The DoE was applied again creating 8 experiments with the limit in values as shown in Table \ref{ExperimentPermutations}. 
Note that the order of the experiment must follow exactly the one given in the table. Although the results will not be the optimal ones, it is important to finish the set of 8 experiments without modifying any of the variables or its range. The results will be analyzed in the following step, taking into consideration whether it is mandatory to readjust them and make another iteration. 

\begin{table}[htbp]
	\centering
	\caption{The first experiment set results.}
	\label{ResultsTable1}
	\scalebox{0.6}{\begin{tabular}{cccccccccccccccccc}
			\toprule
			\textbf{Experiment} & \textbf{A}& \textbf{Value} & \textbf{B} & \textbf{Value} & 
			\textbf{C} &\textbf{Value} & \textbf{AB} & \textbf{Value} & 
			\textbf{AC}& \textbf{Value} & \textbf{BC} &\textbf{Value}& 
			\textbf{Round 1} &\textbf{Round 2} &\textbf{Round 3} & \textbf{Std. Dev.} &\textbf{Average}\\
			\midrule
			1&-&-1&-&-1&-&-1&+&1&+&1&+&1&0.0724 & 0.0808 & 0.0685 & 0.0063 & 0.0739\\
			2&-&-1&-&-1&+&1&+&1&-&-1&-&-1&0.0726 & 0.0811 & 0.0612 & 0.0100 & 0.0716\\
			3&-&-1&+&1&-&-1&-&-1&+&1&-&-1&0.0570 & 0.0748 & 0.0631 & 0.0090 & 0.0650\\
			4&-&-1&+&1&+&1&-&-1&-&-1&+&1&0.0597 & 0.0645 & 0.0664 & 0.0034 & 0.0636\\
			5&+&1&-&-1&-&-1&-&-1&-&-1&+&1&0.1424 & 0.2098 & 0.1703 & 0.0339 & \textbf{0.1741}\\
			6&+&1&-&-1&+&1&-&-1&+&1&-&-1&0.1428 & 0.2112 & 0.1707 & 0.0344 & \textbf{0.1749}\\
			7&+&1&+&1&-&-1&+&1&-&-1&-&-1&0.1292 & 0.1634 & 0.1353 & 0.0182 & 0.1426\\
			8&+&1&+&1&+&1&+&1&+&1&+&1&0.1294 & 0.1645 & 0.1351 & 0.0188 & 0.1430\\
			Effect &   & 0.0901 &   & -0.0201 & 
			  & -0.0006 &   & -0.0116 & 
			  & 0.0012 &  & 0.0001 &   &   &   &   &  \\
			Coefficient &   & 0.0451 &   & -0.0101
			&   & -0.0003 &    & -0.0058
			&  
			& 0.0006 &   & 0.0001
			&   &   &   &   &  \\
			\bottomrule
	\end{tabular}}
\end{table}

After acquiring and processing the traces the ``Round 1'' column in Table \ref{ResultsTable1} was filled in and the highest correlation level is obtained between the stored data and the power traces. It should be noticed that, as mentioned above, the set of 8 experiments has been performed another 2 times (columns ``Round 2'' and ``Round 3'') to ensure that the results are consistent. The coefficients are calculated with the averaged data (column ``Average'').

After applying Eq. (\ref{coef1}) and Eq. (\ref{coef2}) we derived the ``Coefficients'' of the tuning of the parameters and their interactions. The effect that each parameter has can be seen in Fig. \ref{fig:iDOE_1_1_AVG_&_PARETO} (left). Also, in Fig. \ref{fig:iDOE_1_1_AVG_&_PARETO} (right) we show how ``vital few'' variables have more influence over the results than the ``trivial many'' variables. This effect it is also known as the 80/20 rule, stating that, for many events, the 80\% of consequences come from 20\% of causes. This has become a popular maxim in several fields like economics, production, business, computer science, etc.  For instance, Lowell Arthur noted that, in computing, the 20\% of the code has 80\% of the errors. Also, he discovered that the 80\% of a particular piece of software can be written in 20\% of the total allocated time \cite{ParetoExample}. More examples can be found in literature \cite{Pareto}. In this case the \textit{cumulative} line crosses the line of 80\% with the first variable, what basically means that the point of alignment is the only variable which is affecting the results in a significant way (it has almost the 80\% of the influence in the results). At this point, we have one set of the results and hence the factors of Eq. (\ref{doe1}) given by DoE. The parameters A, B, and C are given in Table \ref{ParameterSelection1} so the results given by Eq. (\ref{doe1}) mean that our experiment has better results when the \textbf{Point of alignment} is \textbf{Align at the end} and not when \textbf{LowPass filter} is applied. In this setup the \textbf{Standardization} effect is negligible. 

\begin{dmath}
\label{doe1} DoE = 0.101 + 0.035*A - 0.007*B + 0.0004*C + 0.0002*AB - 0.0003*AC + 0.0003*BC
\end{dmath}

\subsection{Improve}
\label{UC1.IMPRO}

With the information derived in the previous steps, the expert should analyze the results of DoE, interpret them, and decide if they are good enough to implement them in the acquisition setup (the \textit{OK-Criterium} assists in this task). The expert can also dig deeper in his interpretation of the data and propose modifications in the range of some parameter(s), because he/she can derive from the results that the best approach might be keeping a better balance between the different parameters, instead of using the limited values for some of them. In this last case, the effect of the interactions can be very relevant to make a decision. On the other hand, if the decision is to discard the proposed changes, two options are left: go back to Step 3 (Analyze) and perform another iteration of DoE, making changes directly in the definition step based on the gathered knowledge; or perform the side-channel evaluation of the DUT with the baseline setup.

\begin{table}[htbp]
		\centering
		\caption{Working variables and values}
		\label{ParameterSelection2}
		\scalebox{1}{\begin{tabular}{cccc}
				\toprule
				\textbf{ID} & \textbf{Parameter} 	&  \textbf{$1^{st}$ option (-)}& 
				\textbf{ $ 
					2^{nd}$ option (+)}\\
				\midrule
				A & Number of traces & 3000 & 5000\\ 
				B & Windowed resample & false & true\\ 
				C & Standardization & false & true\\
				\bottomrule
		\end{tabular}}
\end{table}

\begin{table}[htbp]
		\centering
		\caption{Results of the three experimental rounds}
		\label{ExperimentRounds2}
		\scalebox{1}{\begin{tabular}{cccccccccc}
				\toprule
				\textbf{Exp} & \textbf{A} & \textbf{B} & \textbf{C} & \textbf{Round 1} & 
				\textbf{Round 2} & \textbf{Round3} & \textbf{Std. Dev.} & \textbf{Average} \\
				\midrule
				1 & 3k & false & false & 0.5754 & 0.5649 & 0.6008 & 0.0185 & \ \textbf{0.5804}\\
				2 & 3k & false & true & 0.5768 & 0.5691 & 0.6059 & 0.0194 & \textbf{0.5839}\\
				3 & 3k & true & false & 0.5665 & 0.5805 & 0.6079 & 0.0210 & \textbf{0.5850}\\
				4 & 3k & true & true & 0.5708 & 0.5855 & 0.6146 & 0.0223 & \textbf{0.5903}\\
				5 & 5k & false & false & 0.5959 & 0.6031 & 0.6276 & 0.0166 & \textbf{0.6089}\\
				6 & 5k & false & true & 0.5990 & 0.6078 & 0.6308 & 0.0164 & \textbf{0.6126}\\
				7 & 5k & true & false & 0.6073 & 0.5978 & 0.6151 & 0.0086 & \textbf{0.6067}\\
				8 & 5k & true & true & 0.6115 & 0.6054 & 0.6178 & 0.0062 & \textbf{0.6116}\\
				\bottomrule
		\end{tabular}}
\end{table}

As it can be noticed in Table \ref{ResultsTable1}, aligning at the end of the operation allows us to find better correlation values (approximately 0.1 larger values). We see some significant correlation spikes in comparison with the correlation obtained in non-leaking parts of the signal, but the correlation level is still too low (only the correlation in Experiments 5 and 6 is barely out of the confidence interval). Thus, we decided to perform another iteration (another DoE) modifying some variables, as shown in Table \ref{ParameterSelection2}. We fix the variable ``point of the alignment'' in its maximum value (align at the end). In other words, we improved the alignment with the focus in the leaking part of the signal. As we obtained better results without applying a lowpass filter, we fix that variable (no lowpass filter) and we add two new variables to analyze (Number of traces, and Compression technique). We keep the variable Standardization to discover whether it can improve the results with the new point of alignment. Repeating the same steps (but with the three new variables) we obtained the results shown in Table \ref{ExperimentRounds2}.
	
\begin{figure}[htbp]
	\centering
	\scalebox{1}{\includegraphics[width=0.9\textwidth]{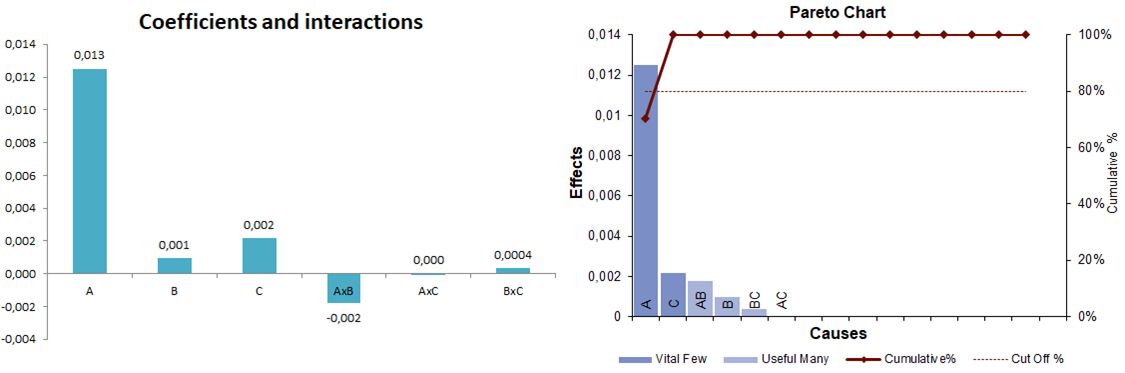}}
	\caption{\label{fig:iDOE_1_2_&_PARETO} Coefficients and interactions (left) and Pareto chart (right)}
\end{figure}

\begin{figure}[htbp]
	\centering
	\scalebox{1}{\includegraphics[width=1\textwidth]{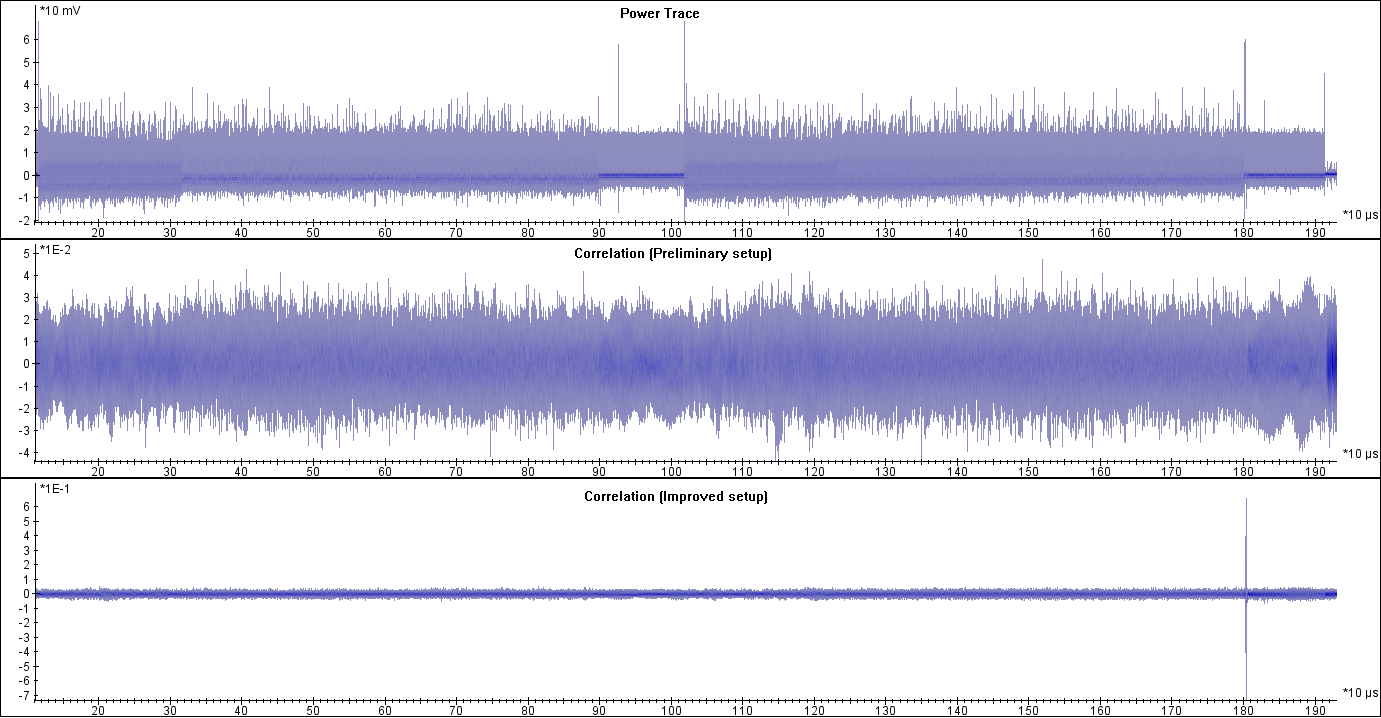}}
	\caption{\label{fig:icorrelationFound} Correlation results}
\end{figure}

These results show that our experiment has better results when the
\textbf{number of power traces} is \textbf{5\,000}, using a compression technique (\textbf{Windowed resample = true}) and using \textbf{Standardization}. Observe that the variable with more effect over the obtained correlation is the number of traces. Also, with this setup the results are slightly better using standardization, contrary to previous setup. It must be mentioned that with this setup the alignment has been improved (due to the results of the previous iteration) and the \textit{OK-criterium} has been accomplished in all the experiments. Fig. \ref{fig:icorrelationFound} shows the differences between the traces taken with the preliminary setup and the traces obtained with the optimized setup. In the bottom chart named Correlation (improved setup) a high correlation spike can be seen at the end of the power trace, indicating the exact place in which the 8-bit value is stored in memory.

\subsection{Control}

Although there is no process running that should be controlled (strictly speaking), the recommendation is: document everything, every step taken, every guess done, every clue discovered, etc. It is the key for having under control all the processes described above, and also for future improvements that can be done for this or other acquisition setups.

\section{Use Case 2: Side-Channel Attack}
\label{UC2}
In the previous section, we have shown how the proposed methodology is used for optimizing the acquisition setup. In this section, we describe how this methodology can be used to optimize an attack scenario. As in the previous section, we first briefly describe the presented use case, and then the 5 DMAIC scheme steps are discussed (except the control step). As the basis of the methodology have been already explained, we focus only on the results. 

\subsection{Use Case description}
\label{UC2.UCD}
The target is an ATmega138P 8-bit microcontroller. Influenced by the result of the previous use case, we want to perform a template attack over a different device in a similar setup. We are storing data (8-bit values) in flash memory using a \textit{memcpy()} operation (in a random address each time). During that operation, we take measurements of the power consumption of the device. As an attacker, our goal is to obtain the exact 8-value loaded in flash memory using template attacks \cite{TA1,TA2,TA3}. Using our customized Six Sigma methodology we have been able to successfully recover the 8-bit value performing template attacks. We performed two different phases: a profiling phase, and an attack phase. In the profiling phase we modeled the side-channel power consumption of the device during the \textit{memcpy()} operation (taking measurements of the device loading random values into the memory), then in the attack phase we were able to guess a (fixed) secret 8-bit value. Since performing a template attack is a complex process with lots of variables involved, we show how our customized Six Sigma methodology can help to optimize the attacking process.

\subsection{Define}
\label{UC2.DEF}

As mentioned above, the main goal of this phase is to define the setup requirements, see Table \ref{setup_goal2}. For us, the main goal is to obtain an 8-bit value loaded into memory by using template attacks. The \textit{OK-criterium} indicates that the correct key guess obtains a rank of 5 or less using template attacks. Table \ref{setup_goal2} shows the setup's requirements.

\begin{table}[htbp]
	\centering
	\caption{Goal and \textit{OK-criterion}}
	\label{setup_goal2}
	\scalebox{1}{\begin{tabular}{ T  E }
			\toprule
			Goal & Successfully obtain the 8 bits loaded into memory using template attacks
			\\
			\hline
			Requirements & \begin{minipage}[t]{0.65\textwidth} \begin{itemize} 
					\item Same setup and device operational parameters fixed 
					and constant in each experiment
					\item Two sets of traces: one with random data (profiling 
					phase) and other with constant data (attacking phase)
					\item Oscilloscope: LeCroy Waverunner 9104 
					\item SW for the acquisition, signal processing and data 
					analysis 
					\item Current probe
			\end{itemize} \end{minipage}\\ 
			\hline
			\textit{OK-criterion} & The correct candidate obtains a rank of 5 or less using template attacks\\
			\bottomrule
	\end{tabular}}
\end{table}

\subsection{Measure}
\label{UC2.MEAS}

\begin{table}[htbp]
	\centering
	\caption{Defined variables}
	\label{ParameterTable2}
	\scalebox{1}{\begin{tabular}{cLDLL}
			\toprule
			\textbf{Rank} & \textbf{Parameter} & \textbf{Description} & \textbf{Range} & \textbf{Fixed Value}\\
			\midrule
			1 & Standard-ization &
			Removing the mean of the data set can help to improve the results. &
			Yes vs No &
			\\
			2 & Points of Interest Nº & 
			The selected POI will affect the templates and therefore the attack. We must select an optimal number of points of interest. & 
			1 vs 3 &
			\\
			3 & LowPass Filter (SW) &
			To filter the signal or not would affect the quality of the collected traces and the leakage. We should eliminate high frequency noise but without destroying the leakage &
			With SW filter vs Without SW filter &
			\\
			4 & Nº Traces Profiling &
			Number of processed traces used for the profiling phase of the template attack & 
			1k vs 100k &
			\textbf{20k}\\
			5 & Nº Traces Attack &
			Number of processed traces used for the attacking phase of the template attack & 
			1k vs 10k &
			\textbf{1k}\\
			6 & Alignment & 
			When we align, we can choose different points as a reference. &
			Start vs End &
			\textbf{Start} \\
			7 & POI selection function	&
			We can use different functions to select the points of interest of the traces (SOST~\cite{gierlichs2006templates}, SOSD~\cite{gierlichs2006templates}, SNR~\cite{BlueBook}, CORRELATION~\cite{james2014stat} &
			SOST vs SNR &
			\textbf{SOST}\\
			8 & Compression technique (Windowed resample) &
			A compression technique can be used to reduce the dataset size and to improve the leakage (close points carry very similar information and noise can be reduced). Conversely, in some cases leakage can be destroyed. &
			With compression vs without compression &
			\textbf{No compression}\\
			9 & Sampling Frequency &
			A high sampling frequency will improve the quality of the traces but also increase the data size &
			100MHz vs 1GHz &
			\textbf{1 GHz} \\
			\bottomrule
	\end{tabular}}
\end{table}

With the baseline setup shown in Table~\ref{ToolList2}, a round of 8-bit random values are sent to the DUT by the computer through the serial port, while the oscilloscope is triggered with a GPIO controlled by the ATmega platform, just after finishing the communication, when the internal computation in the DUT is supposed to start.
We set the range of the system's variables, and select three of them to check the effect they have in the success of the attack. Table \ref{ParameterTable2} presents the variables considered for this experiment. 

\begin{table}[!h]
	\centering
	\caption{Tools list and brief description.}
	\label{ToolList2}
	\scalebox{1}{\begin{tabular}{TE}
			\toprule
			LeCroy Waverunner 9104 Oscilloscope & 
			\begin{minipage}[t]{0.65\textwidth} \begin{itemize} 
					\item 2 channels (Power consumption \& triggering)  
					\item 20 GS/s (Single sample capture mode)
				\end{itemize}
			\end{minipage}\\ 
			\hline
			The DUT & \begin{minipage}[t]{0.65\textwidth} \begin{itemize} 
					\item Generates trigger trough GPIO
					\item Power source: continuous power supply
					\item Communicates with PC via serial por
			\end{itemize} \end{minipage}\\ 
			\hline
			PC  & \begin{minipage}[t]{0.65\textwidth} \begin{itemize} 
					\item Communicates with the ATmega138P 
					\item Controls the oscilloscope
			\end{itemize} \end{minipage}\\ 
			\hline
			Current Probe  & \begin{minipage}[t]{0.65\textwidth} \begin{itemize} 
					\item Tektronix CT1 current probe
					\item Conected in series with the DUT power line
					\item Measure the power consumption
			\end{itemize} \end{minipage}\\ 
			\bottomrule
	\end{tabular}}
\end{table}

\subsection{Analyze}
\label{UC2.ANA}

\begin{figure}[htbp]
	\centering
	\scalebox{1}{\includegraphics[width=1\textwidth]{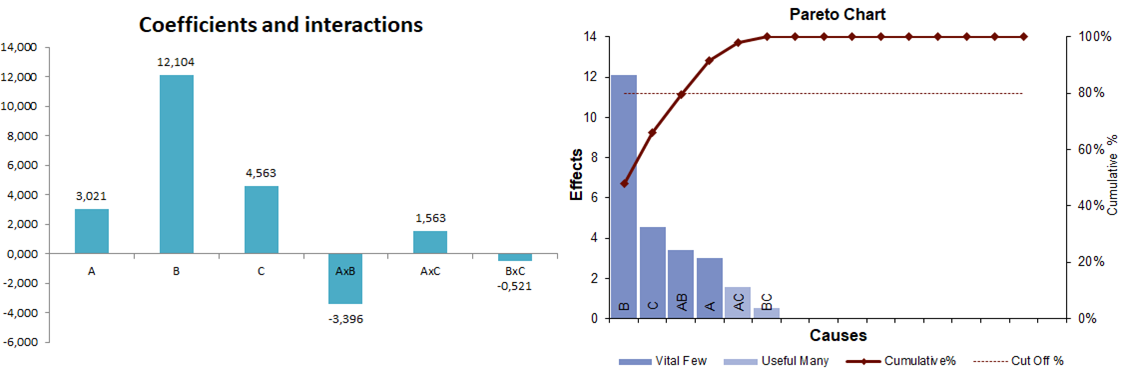}}
	\caption{\label{fig:iDOE_2_1_&_PARETO} Coefficients and interactions (left) and Pareto chart (right)}
\end{figure}

Table \ref{ParameterSelection3} shows the three top variables to use in the DoE, creating 8 experiments. For each experiment 10 k traces of random data were captured for the profiling phase and 1k traces of constant data were captured for the attack phase. The experiments and their results are given in Table \ref{ExperimentRounds3}. The parameters A, B, and C are given in Fig.~\ref{fig:iDOE_2_1_&_PARETO} (left) so the outcomes mean that our experiment has better results when the \textbf{Number of Points of Interest} is \textbf{3}, and when we apply a \textbf{LowPass filter}, and \textbf{Standardization}. We can see that the number of POI is the variable with more effect in the results, but in this case variables A and C have also a significant effect (see Fig.~\ref{fig:iDOE_2_1_&_PARETO} (right)).

\begin{table}[htbp]
		\centering
		\caption{Working variables and values}
		\label{ParameterSelection3}
		\scalebox{1}{\begin{tabular}{cccc}
				\toprule
				\textbf{ID} & \textbf{Parameter} 	&  \textbf{$1^{st}$ option (-)}& 
				\textbf{ $ 2^{nd}$ option (+)}\\
				\midrule
				A & Standardization & false & true \\ 
				B & Nº POI & 1 & 3\\ 
				C & Lowpass Filter (SW) & false & true\\
				\bottomrule
		\end{tabular}}
\vspace{0.5cm}
		\centering
		\caption{Results of the three experimental rounds}
		\label{ExperimentRounds3}
		\scalebox{0.8}{\begin{tabular}{ccccccccccccc}
				\toprule
				\textbf{Exp} & \textbf{A} & \textbf{B} & \textbf{C} & \textbf{Round 1} & 
				\textbf{Round 2} & \textbf{Round 3} & \textbf{Round 4} & 
				\textbf{Round 5} & \textbf{Round 6} & \textbf{Std. Dev.} & \textbf{Average} \\
				\midrule
				1 & false & 1 & false & 46 & 63 & 42 & 27 & 47 & 27 & 13.65 & 42\\
				2 & false & 1 & true & 24 & 71 & 27 & 37 & 41 & 22 & 18.25 & 37\\
				3 & false & 3 & false & 8 & 17 & 7 & 2 & 4 & 34 & 11.95 & 12\\
				4 & false & 3 & true & 5 & 13 & 3 & 2 & 3 & 4 & 4.05 & 5\\
				5 & true & 1 & false & 79 & 23 & 26 & 24 & 31 & 23 & 22.09 & 34.33\\
				6 & true & 1 & true & 41 & 1 & 21 & 19 & 13 & 19 & 13.02 & 19\\
				7 & true & 3 & false & 23 & 16 & 2 & 5 & 5 & 32 & 11.96 & 13,83\\
				8 & true & 3 & true & 9 & 9 & 5 & 1 & 2 & 2 & 3.61 & \textbf{4,67}\\
				\bottomrule
		\end{tabular}}
\end{table}

\subsection{Improve}
\label{UC2.IMPRO}

We reached the \textit{OK-criterium} only in Experiment 8. Since we consider it a very poor result, we moved forward to increase the success rate and we performed another DoE iteration. Then, we redefined the 3 variables as shown in Table \ref{ParameterSelection4}. 

\begin{table}[!h]
		\centering
		\caption{Working variables and values}
		\label{ParameterSelection4}
		\scalebox{1}{\begin{tabular}{cccc}
				\toprule
				\textbf{ID} & \textbf{Parameter} &  \textbf{$1^{st}$ option (-)}& 
				\textbf{ $ 2^{nd}$ option (+)}\\
				\midrule
				A & Strength of the lowpass filter & 1 & 10 \\ 
				B & Nº POI & 3 & 5\\ 
				C & Nº Traces Profiling phase & 5 k & 15 k\\
				\bottomrule
		\end{tabular}}
\vspace{0.5cm}
		\centering
		\caption{Results of the three experimental rounds}
		\label{ExperimentRounds4}	
		\scalebox{0.9}{\begin{tabular}{ccccccccccc}
				\toprule
				\textbf{Exp} & \textbf{A} & \textbf{B} & \textbf{C} & \textbf{Round 1} & 
				\textbf{Round 2} & \textbf{Round 3} & \textbf{Round 4} & \textbf{Std. Dev.} & \textbf{Average} \\
	    \midrule
		1 & 1 & 3 & 5k & 4 & 22 & 7 & 102 & 46.18 & 33.75\\
		2 & 1 & 3 & 15k & 3 & 2 & 4 & 1 & 1.29 & \textbf{2.5}\\
		3 & 1 & 5 & 5k & 3 & 26 & 4 & 112 & 51.60 & 36.25\\
		4 & 1 & 5 & 15k & 1 & 4 & 1 & 1 & 1.50 & \textbf{1.75}\\
		5 & 10 & 3 & 5k & 1 & 15 & 4 & 30 & 13.13 & 12.5\\
		6 & 10 & 3 & 15k & 2 & 1 & 2 & 1 & 0.58 & \textbf{1.5}\\
		7 & 10 & 5 & 5k & 3 & 15 & 2 & 32 & 13.98 & 13\\
		8 & 10 & 5 & 15k & 2 & 1 & 2 & 1 & 0.58 & \textbf{1.5}\\
		\bottomrule
		\end{tabular}}
\end{table}

\begin{figure}[htbp]
	\centering
	\scalebox{1}{\includegraphics[width=1\textwidth]{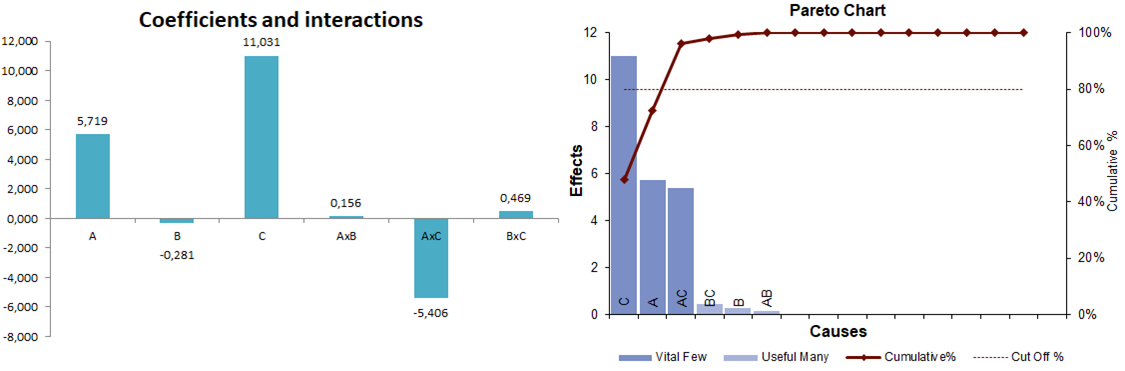}}
	\caption{\label{fig:iDOE_2_2_&_PARETO} Coefficients and interactions (left) and Pareto chart (right)}
\end{figure}

We modified the variable \textbf{Lowpass Filter} to \textbf{Strength of the lowpass filter} in order to obtain the proper value for characterizing the filter. Also, we wanted to know if adding 
more POI could improve the attack so we modified the range (from 3 to 5). Moreover, we decided to check if by using more traces one could improve the attack significantly, so we added the parameter \textbf{Nº Traces Profiling phase}. Table 
\ref{ExperimentRounds4} shows the experiment definition and the results. Note that the \textit{OK-criterium} was 
reached in the half of the experiments (experiments with 15 k traces for profiling phase), so we can consider our main goal reached. Again, the parameters A, B, and C are given in Figure \ref{fig:iDOE_2_2_&_PARETO} (left), and this time the outcomes show that our experiment has better results whit \textbf{15 k Traces for Profiling phase} and \textbf{Standardization}. Note that adding more POI does not improve the results in a significant way, so it is not worth increasing the computational effort required by adding more POI. In Figure \ref{fig:iDOE_2_2_&_PARETO} (right), we see that the variables with more impact to the results are the \textbf{Number of Traces for Profiling phase} (47.8\%) and \textbf{Standardization} (24.8\%).

\section{Use Case 3: Leakage Assessment (Statistical tests)}
\label{UC3}
The following use case shows how our customized Six Sigma methodology can be used in a side-channel evaluation scenario to optimize the leakage assessment process with traditional statistical tests (e.g. $t$-test and $\mathcal{X}^2$-test). We first present a short overview of the current leakage assessment methodologies. Then we describe our use case and the 5 DMAIC scheme steps. 

\subsection{Leakage Assessment Methodologies: why is it necessary?}
\label{UC3.LAO}

The need for integration and validation of countermeasures against side-channel attacks on embedded devices earned quite some attention in recent years. Current certification process like EMVCo\cite{EMVCospringer} or Common Criteria (CC)\cite{CCspringer} evaluate the robustness of a DUT by directly attacking it with different side-channels techniques (e.g. differential power analysis~\cite{kocher1999dpa}, correlation power analysis~\cite{brier2004cpa}, mutual information analysis~\cite{Gierlichs2007mutual,lejla2011mia}, template attacks~\cite{TA1,TA2,TA3}, deep learning-based attacks~\cite{Maghrebi2016BreakingCI,CNNOV,masure2019comprehensive}, etc.). Nevertheless, the increasing amount of attacks, possible algorithms, distinguishers and models make this kind of evaluation infeasible, in a low-cost and efficient manner, specially for less experienced evaluators.

To overcome those drawbacks, several leakage assessment techniques have arisen in recent times with the purpose of determining, whether a device leaks information through side channels in a quick and simple way. These tests alone are not enough for evaluating a device against SCA, since they do not quantify the leakage or give any clue about its exploitability, but are good for a preliminary (``black-box'') evaluation. The most popular one is the Test Vector Leakage Assessment (TVLA) methodology by Cryptography Research (CRI) \cite{NISTcri,NISTcri&risc,Becker2013TestVL}. The approach is to use a statistical test (commonly Welch's $t$-test or Pearson's $\mathcal{X}^2$-test \cite{Moradi2018}) to distinguish whether two sets of data (e.g.random vs fixed) are significantly different. However, performing TVLA properly is not a trivial task, since there are lots of possible choices and variables to tune. Thus, we propose to use our methodology to improve the leakage assessment process by helping to tune its parameters.

\subsection{Use Case description}
\label{UC3.UCD}
Our target is an STM32F417 32-bit microcontroller implementing a software AES-128 implementation \cite{daemen02,daemen01springer,Bertoni:2002:ESI:648255.752733}. As evaluators, our goal is to detect leakage with the least amount of resources possible (fast and efficient).
Essentially, what we want to do is to prove that there exists a dependency between the data processed by the cryptographic algorithm and the power consumption. Commonly, the DUT is fed with two differentiated types of data (fixed vs fixed, fixed vs random, semi-fixed vs random, etc.) and the evaluator tries to confirm that there exists a significant statistical difference between both sets of traces.

As mentioned above, the TVLA is the main statistical tool used in side-channel leakage detection since it is fast and versatile. However, it can also bring up false positives~\cite{Standaert2017HowT}. For this reason, in real scenario evaluations, a semi-fixed vs random test is performed (instead of a fixed vs random test), and also for the same reason new suggested evaluation techniques are published~\cite{SFMRT,FastLA,Moradi2018,DBLP:journals/iacr/WegenerM019}.

In other words, instead of feeding the cryptographic device with fixed values, we generate specific test vectors which force a certain intermediate value (or its Hamming Weight) to remain constant. Then we compare this power consumption, with the one generated when the device is encrypting fully random values. The usage of semi-fixed vectors makes the TVLA test more reliable, but it increases the complexity of the evaluation with more parameters to tune. For this reason, our customized Six Sigma fits also perfectly in this use case.

\subsection{Define}
\label{UC3.DEF}

Table \ref{setup_goal3} shows the main goal and \textit{OK-criterion} of the phase. Here, the main goal is to detect leakage on an AES-128 bit software implementation running on an STM32F417 microcontroller. Thus, the \textit{OK-Criterium} in this use case is to obtain a $p$-value greater than $ p = 10^{-5}$, the common threshold in leakage assessment evaluation. This value is equivalent to the widely used threshold of 4.5 in $t$-test. After the acquisition, a statistical test ($t$-test or $\mathcal{X}^2$-test) over the samples is run. It is important to perform the statistical test over two different sets of data (taken with the same setup), to confirm that the leakage appears in both sets of data at the same time.

\begin{table}[!t]
	\centering
	\caption{Goal and \textit{OK-criterion}}
	\label{setup_goal3}
	\scalebox{1}{\begin{tabular}{ T  E }
		\toprule
		Goal & Detect leakage on an AES-128 bit software implementation
		\\
		\hline
		Requirements & \begin{minipage}[t]{0.65\textwidth} \begin{itemize} 
			\item Same setup and device operational parameters fixed and constant in each experiment
			\item Two sets of traces: one with random data and other with semi-constant data
			\item Oscilloscope: LeCroy Waverunner 9104 
			\item SW for the acquisition, signal processing and data analysis 
			\item Current probe
		\end{itemize} \end{minipage}\\ 
		\hline
		\textit{OK-criterion} & Obtain a $p$-value greater than $ 10^{-5} $ at the 
		same time in two different sets of samples \\
		\bottomrule
	\end{tabular}}
\end{table}

\subsection{Measure}
\label{UC3.MEAS}

\begin{table}[!b]
	\centering
	\caption{Tools list and brief description.}
	\label{ToolList3}
	\scalebox{1}{\begin{tabular}{ T  E }
		\toprule
		LeCroy Waverunner 9104 Oscilloscope & 
		\begin{minipage}[t]{0.65\textwidth} \begin{itemize} 
				\item 2 channels (Power consumption \& triggering) 
                \item 20 GS/s (Single sample capture mode)
			\end{itemize}
		\end{minipage}\\ 
		\hline
		The DUT  & \begin{minipage}[t]{0.65\textwidth} \begin{itemize} 
			\item Generates trigger trough GPIO
			\item Power supplied through external batteries
			\item Communicates with PC via serial port
		\end{itemize} \end{minipage}\\ 
		\hline
		PC  & \begin{minipage}[t]{0.65\textwidth} \begin{itemize} 
			\item Communicates with the STM32F417 microcontroller
			\item Controls the oscilloscope
			\item Generates random data to encrypt
		\end{itemize} \end{minipage}\\ 
		\hline
		Current Probe  & \begin{minipage}[t]{0.65\textwidth} \begin{itemize} 
			\item Tektronix CT1 current probe
			\item Connected in series with the DUT power line
			\item Measure the power consumption
		\end{itemize} \end{minipage}\\ 
		\bottomrule
	\end{tabular}}
\end{table}

\begin{table}[!p]
	\centering
	\caption{Defined variables}
	\label{ParameterTable3}
	\scalebox{1}{\begin{tabular}{cLDLL}
		\toprule
		\textbf{Rank} & \textbf{Parameter} & \textbf{Description} & \textbf{Range} & \textbf{Fixed Value}\\
		\midrule
		1 & Intermediate value 	&
		Semi-fixed traces are generated such that certain intermediate value Hamming Weight is always between a range. The intermediate value targeted can have an influence in the leakage detection.	&
		SubBytes vs AddRoundKey	&
		\\
		2 & HW range & 
		The HW range of the generated semi-fixed traces can have an influence in the leakage detection &
		40-60 vs 80-100	&
		\\
		3 & Statistical test &
		Although the $t$-test is considered to be the main statistical tool in side channel detection, recently the $\mathcal{X}^2$-test for the same purpose has been proposed. The usage of one statistical tool or another can have clear influence in the obtained results. &
		$t$-test vs $\mathcal{X}^2$-test &
		\\
		4 & Test Vector &
		The nature of the test vector may affect the results. We could use the classical fixed vs random approach or we can generate specific vectors which force the HW of one intermediate value &
		"Fixed vs random'' VS ``Semi-Fixed vs random'' & 
		\textbf{Semi-Fixed vs random}\\
	    5 & Number of traces &
		The number of traces affects the confidence of the results. In this case in each experiment we need two sets of traces taken with the same setup & 
		1k vs 100k &
		\textbf{5k}\\
		6 & Standard-ization &
		Removing the mean of the data set can help to improve the results. &
		Yes vs No & 
		\textbf{With standardization}\\
		7 & Compression (Windowed resample) &
			A compression technique can be used to reduce the data set size and improve the leakage (close points carry similar information and noise can be reduced). Conversely, in some cases leakage can be destroyed. &
			With compression vs without compression &
		\textbf{No compression}\\
		8 & Alignment & 
			When we align, we can choose different points as a reference. &
			Start vs End &
			\textbf{Start} \\
		9 & LowPass Filter (SW) &
		To filter the signal or not would affect the quality of the collected traces and the leakage. We should eliminate high frequency noise but without destroying the leakage &
		With SW filter vs Without SW filter &
		\textbf{Without LowPass filter}\\
		10 & Sampling Frequency &
		A high sampling frequency will improve the quality of the traces but also increase the data size. &
		50MHz vs 20GHz &
		\textbf{1 GHz} \\
		\bottomrule
	\end{tabular}}
\end{table}

With the baseline setup parameters as in Table \ref{ToolList3}, a round of random or semi-fixed 128-bit values (randomly interleaved) are sent to the DUT by the computer trough the serial port, while the oscilloscope is triggered with a GPIO controlled by the DUT, just after finishing the communication (when the internal computation in the DUT is supposed to begin). We evaluate the system's variables, and select three of them to check the effect they have in the success of the attack (Table \ref{ParameterTable3}). In this case, we have selected the three variables shown in Table \ref{ParameterSelectionTVLA}. We want to know which fixed intermediate value produces higher $p$-values, so we select variable A (Intermediate Value). Also, we want to select the proper Hamming Weight range for the semi-fixed set (for the selected intermediate value). Finally, we want to know which statistical test ($t$-test or $\mathcal{X}^2$-test) works better in this evaluation.

\subsection{Analysis}
\label{UC3.ANA}

In this step, the DoE was applied on the 3 aforementioned variables creating 8 experiments. The experiments and their results are presented in Table \ref{ExperimentRoundsTVLA}. The parameters A, B, and C are given in Figure \ref{fig:iDOE_3_1_&_PARETO}, and the outcomes mean that our experiment has better results when the \textbf{Intermediate value} is \textbf{SubBytes}, the \textbf{HW range} is \textbf{80-100} and we use the \textbf{$\mathcal{X}^2$-test} for the experiment. 

\begin{table}[htbp]
	\centering
	\caption{Working variables and values}
	\label{ParameterSelectionTVLA}
	\scalebox{1}{\begin{tabular}{cccc}
			\toprule
			\textbf{ID} & \textbf{Parameter} &  \textbf{$1^{st}$ option (-)}& 
			\textbf{ $2^{nd}$ option (+)}\\
			\midrule
			A & Intermediate value & SubBytes & AddRoundKey \\ 
			B & HW range & 40-60 & 80-100\\ 
			C & Statistical test & $t$-test & $\mathcal{X}^2$-test\\
			\bottomrule
	\end{tabular}}
\vspace{0.5cm}
	\centering
	\caption{Results of the six experimental rounds}
	\label{ExperimentRoundsTVLA}
	\scalebox{0.9}{\begin{tabular}{cccccccccc}
		\toprule
		\textbf{Exp} & \textbf{A} & \textbf{B} & \textbf{C} & \textbf{Round 1} & 
		\textbf{Round 2} & \textbf{Round 3} & \textbf{Std. Dev.} & \textbf{Average} \\
		\midrule
		1 & SubBytes & 40-60 & $t$-test & 14.105 & 12.365 & 11.250 & 1.439 & \textbf{12.573}\\
		2 & SubBytes & 40-60 & $\mathcal{X}^2$-test & 9.889 & 13.256 & 11.143 & 1.702 & \textbf{11.429}\\
		3 & SubBytes & 80-100 & $t$-test & 64.930 & 51.837 & 51.198 & 7.750 & \textbf{55.988}\\
		4 & SubBytes & 80-100 & $\mathcal{X}^2$-test & 59.621 & 53.409 & 55.266 & 3.188 & \textbf{56.099}\\
		5 & AddRoundKey & 40-60 & $t$-test & 3.886 & 3.907 & 3.620 & 0.160 & 3.804\\
		6 & AddRoundKey & 40-60 & $\mathcal{X}^2$-test & 6.802 & 6.025 & 6.182 & 0.411 & \textbf{6.337}\\
		7 & AddRoundKey & 80-100 & $t$-test & 5.790 & 5.641 & 4.613 & 0.641 & \textbf{5.348}\\
		8 & AddRoundKey & 80-100 & $\mathcal{X}^2$-test & 48.948 & 41.981 & 38.256 & 5.427 & \textbf{43.061}\\
		\bottomrule
	\end{tabular}}
\end{table}

\subsection{Improve}
\label{UC3.IMPRO}

\begin{figure}[htbp]
	\centering
	\scalebox{1}{\includegraphics[width=1\textwidth]{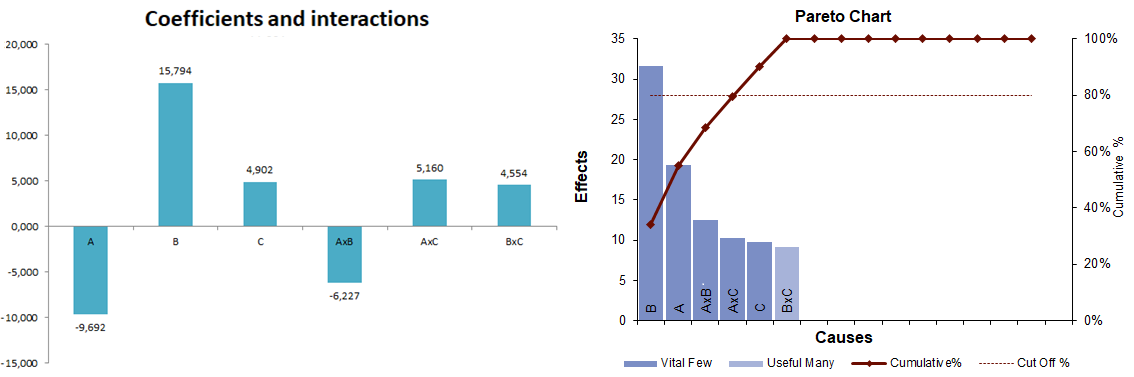}}
	\caption{\label{fig:iDOE_3_1_&_PARETO} Coefficients and interactions (left) and Pareto chart (right)}
\vspace{0.5cm}
	\centering
	\scalebox{1}{\includegraphics[width=1\textwidth]{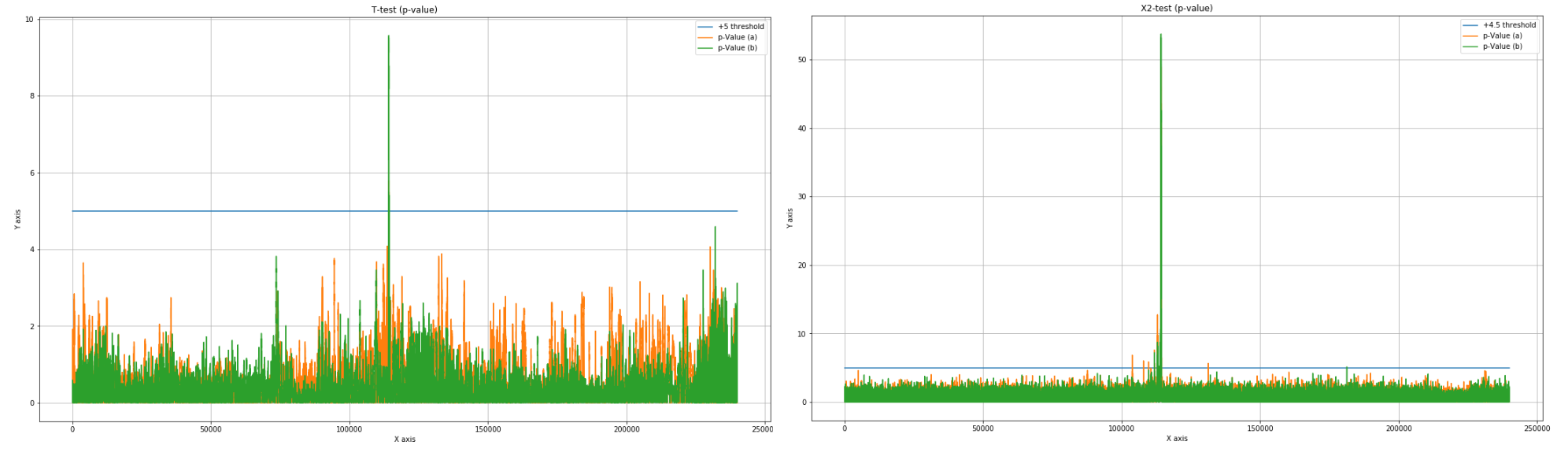}}
	\caption{\label{fig:ittest} Leakages of AddRoundKey intermediate value: $t$-test vs $\mathcal{X}^2$-test (Experiments 7 and 8)}
\end{figure}

Since the \textit{OK-criterium} has been reached in most of the experiments, it is not necessary to perform another iteration of the DoE. We conclude that the device leaks information through its power consumption. From the results of the previous step, it is noticeable that the output of SubBytes is leakier than the one of AddRoundKey. Also, it can be seen that the higher Hamming Weight value we fix for the semi-fixed dataset, the larger statistical differences are observed in the traces. As it is observed in the Pareto chart (Figure \ref{fig:iDOE_3_1_&_PARETO}) there are variables with more effect than others, but the particularity is that the three of them have a strong influence on the results. Conversely, it is known that $\mathcal{X}^2$-test gives better results in the cases where the leakage has multivariate behaviour. Nevertheless, its authors confirmed that the technique is feasible also in univariate cases (our case)\cite{Moradi2018}. In our experiment while using the $t$-test, we were not able to exceed the threshold by fixing the HW of the AddRoundKey operation with 5000 traces. Surprisingly, using $\mathcal{X}^2$-test we were able to find leakage with the same number of traces. In other words, in our experiments the $\mathcal{X}^2$-test was more sensitive and allowed us to detect differences better than $t$-test. Thus, although $\mathcal{X}^2$-test requires more computational effort, using it improves the results in particular cases (as it can be seen in Figure \ref{fig:ittest}).

\section{Use Case 4: Leakage Assessment (Deep Learning)}
\label{UC4}

Recently, a new leakage assessment method based on deep learning has been proposed~\cite{DBLP:journals/iacr/WegenerM019}. Their idea is to train a neural network that works as a classifier over the two sets of data (e.g. random vs fixed). If the neural network is able to distinguish between these two sets of data, it can be assumed that they are statistically different with a certain probability. Additionally, dealing with pre-processing problems like misalignment is becoming less problematic, if at all needed. Conversely, the inclusion of deep neural networks to the leakage assessment adds complexity to to the problem. In this section we show the viability of our customized Six Sigma methodology in deep leaning leakage assessment, helping to discern which is the best setup for this purpose.

\subsection{Use Case description}
For this case, the target, preliminary setup, and main goal are the same as in the previous use case. The neural network has the same architecture as defined in the original paper \cite{DBLP:journals/iacr/WegenerM019}. The set of ``semi-fixed vs random'' power traces is divided into training and validation set. Then from the training set, we took 1k and 3k traces to train the model, and from the validation set (10k traces) we compute a binomial test (as is suggested in the original paper) to obtain a probability (p-value) that indicates whether there is a significant statistical difference between the two set of traces (``semi-fixed vs random''). 

Note that, the difference between the original experiment from the paper and this one, is that we are using a microcontroller instead of an FPGA platform, and a different encryption algorithm (AES-128 instead of PRESENT \cite{Bogdanov:2007:PUB:1421964.1422007}).

\subsection{Define}
\label{UC4.DEF}

As mentioned above, the setup requirements, main goal and \textit{OK-criterium} are the same as in the previous use case (Sect.~\ref{UC3.DEF}) and can be found in Table \ref{setup_goal3}. 

\subsection{Measure}
\label{UC4.MEAS}
The baseline setup (Table \ref{ToolList3}) is also the same as in the previous use case in Sect.~\ref{UC3.MEAS}. However, the leakage assessment method is completely different, and thus, the defined variables too. It should be noted that general signal processing variables that are (most of the time) considered in this step (Lowpass filtering, number of traces, compression techniques, etc.) were not applied, since deep learning approaches in SCA are often advocated by the needlessness of pre-processing techniques. However, the authors in \cite{DBLP:journals/iacr/WegenerM019} propose \textit{standardizing} the training and validation sets to obtain a homogeneous range between all input points and weights. Hence, we will consider \textit{standardizing} as a variable in our experiments to check whether this method improves the results of our use case. Other considered variables are the number of traces used for training and validation, and the number of samples per trace in each one of them. Table \ref{ParameterTable4} shows the defined variables of this step. 

\begin{table}[htbp]
	\centering
	\caption{Defined variables}
	\label{ParameterTable4}
	\scalebox{1}{\begin{tabular}{cLDLL}
		\toprule
		\textbf{Rank} & \textbf{Parameter} & \textbf{Description} & \textbf{Range} & \textbf{Fixed Value}\\
		\midrule
		1 & Standardizing  	&
		To reach a homogeneous range between all input points and weights, enabling efficient training	&
		No vs Yes &
		\\
		2 & Nº Samples (per trace) & 
		The number of samples will have an influence in the number of neurons of the neural network, which could have an impact in the results. &
		2500 vs 5000	&
		\\
		3 & Nº Traces (Training) &
		The number of traces used in the training phase may affect to the obtained p-values. &
		1000 vs 3000 &
		\\
		4 & Nº Traces (Validation) &
		The number of traces used in the validation phase may affect to the obtained p-values. &
		1k vs 100k & \textbf{10k}
		\\
		5 & Neural Network Architecture &
		The architecture of the neural network could affect to the obtained probabilities. We can follow the architecture proposed in \cite{DBLP:journals/iacr/WegenerM019} or design our own. &
		Neural Network Sequential Model vs CNN & \textbf{Neural Network Sequential Model}
		\\
		\bottomrule
	\end{tabular}}
\vspace{0.5cm}
		\centering
		\caption{Working variables and values}
		\label{ParameterSelectionDL}
		\scalebox{1}{\begin{tabular}{cccc}
				\toprule
				\textbf{ID} & \textbf{Parameter} &  \textbf{$1^{st}$ option (-)}& 
				\textbf{ $ 
					2^{nd}$ option (+)}\\
				\midrule
				A & Standardizing & No Standardizing & Standardizing \\ 
				B & Nº Samples (per trace) &  2500 & 5000\\ 
				C & Nº Traces (Training) & 1000 & 3000\\
				\bottomrule
		\end{tabular}}
\vspace{0.5cm}
	\centering
	\caption{Results of the six experimental rounds}
	\label{ExperimentRoundsDL}
	\scalebox{1}{\begin{tabular}{cccccccc}
		\toprule
		\textbf{Exp} & \textbf{A} & \textbf{B} & \textbf{C} & \textbf{Round 1} & 
		\textbf{Round 2} & \textbf{Std. Dev.} & \textbf{Average} \\
		\midrule
		1 & No Standardizing & 2k5 & 1k & 299.25 & 415.13 & 81.940 & \textbf{357.19}\\
		2 & No Standardizing & 2k5 & 3k & 910.65 & 956.40 & 32.350 & \textbf{933.53}\\
		3 & No Standardizing & 5k & 1k & 244.82 & 323.01 & 55.289 & \textbf{283.91}\\
		4 & No Standardizing & 5k & 3k &  763.51 & 910.65 & 104.044 & \textbf{837.08} \\
		5 & Standardizing & 2k5 & 1k & 208.35 & 206.70 & 1.167 & \textbf{207.52}\\
		6 & Standardizing & 2k5 & 3k & 667.50 & 729.90 & 44.123 & \textbf{698.70}\\
		7 & Standardizing & 5k & 1k & 148.04 & 169.00 & 14.821 & \textbf{158.52}\\
		8 & Standardizing & 5k & 3k & 638.05 & 686.51 & 34.266 & \textbf{662.28}\\
		\bottomrule
	\end{tabular}}
\end{table}

\subsection{Analysis}
\label{UC4.ANA}

The DoE was applied on the three variables shown in Table \ref{ParameterSelectionDL} creating 8 experiments. The experiments and their result are represented in the Table \ref{ExperimentRoundsDL}. Figure \ref{fig:iDOE_3_1_&_PARETO} shows parameters A, B, and C. The outcomes mean that our experiment has better results with \textbf{No Standardizing}, using \textbf{2500 samples per trace} and \textbf{3k traces for training}.

\begin{figure}[htbp]
	\centering
	\scalebox{1}{\includegraphics[width=1\textwidth]{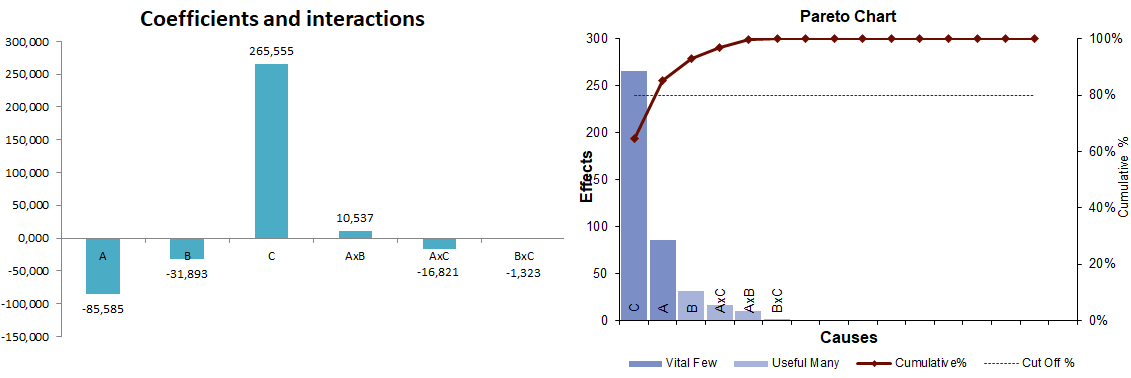}}
	\caption{\label{fig:iDOE_4_1_&_PARETO} Coefficients and interactions (left) and Pareto chart (right)}
\end{figure}

\subsection{Improve}
\label{UC4.IMPRO}

Since the \textit{OK-criterium} has been reached in all the experiments, it is not necessary to perform another iteration of the DoE. From the results of the previous step, we conclude that the number of training traces is the parameter which has more effect on the obtained p-value, as it can be seen in the Pareto Chart (Figure \ref{fig:iDOE_4_1_&_PARETO} (right)). Note that, applying the \textit{standardizing} technique does not improve the results. In this particular case, comparing the same experiments with and without standardizing the traces, we obtain slightly better results without standardizing as pre-processing. However, it is not necessarily conclusive that the pre-processing is not required.

\section{Conclusion}
\label{CONCLUSION}
Our results when using this customized Six Sigma methodology, demonstrate the suitability of this method for improving the SCA process in its different stages; from the basis of the process (which is improving the quality of the acquired side-channel measurement) to the performance of any kind of side-channel attack or leakage assessment technique. Moreover, we have shown how our Six Sigma methodology can reduce the uncertainty associated with the SCA, helping technicians to interpret the results and discover root causes of the phenomena occurred during the process. During the process, the evaluator identifies the parameters that have more influence in the results of a certain experiment and is able to adjust them to an optimal value. 

The methodology steps proposed are simple, methodical and very helpful when dealing with security evaluations. This approach can be helpful to any researcher or security evaluator in a lab; it allows technicians without a deep knowledge of all the basics involved in these methods, to implement and interpret side-channel evaluations properly. The methodology can also be used by experts when dealing with new tasks (e.g. regarding to new devices, attacks or leakage assessment methods), as in such a case methodology could guide the evaluator to find the best set of variables and speed-up the evaluation process.

\bibliographystyle{splncs04}
\bibliography{bib}

\end{document}